\documentclass[aps,prl,reprint,amsmath,amssymb]{revtex4-1}
\usepackage[pdftex,colorlinks]{hyperref}
\usepackage{mathrsfs}
\usepackage{times}

\let\de=\partial
\newcommand\dd{\text{d}}
\newcommand\im{\text{i}}
\newcommand\La{\mathscr{L}}
\newcommand\gr[1]{\mathrm{#1}}% font for group names such as SU(2)

\renewcommand\aa{a}% a^i_A coefficient
\newcommand\bb{b}% b_A coefficient
\newcommand\cc{\bar b}% c_A coefficient
\newcommand\ff{g}% f_{AB} coefficient
\newcommand\g{\bar g}% g_{AB} coefficient
\newcommand\hh{c}% h^i_{AB} coefficient
\newcommand\jj{d}% j_{Ai} coefficient
\newcommand\kk{e}% k^B_{Ai} coefficient
\newcommand\str{f}% \ell^k_{ij} coefficient

\newcommand\LL{\amalg}% \Lambda matrix
\newcommand\XX{\Pi}% \Xi matrix
\DeclareMathOperator{\Co}{\widehat{ch}}
\DeclareMathOperator{\Si}{\widehat{sh}}

\allowdisplaybreaks
\begin{document}

\title{Geometry of Multi-Flavor Galileon-Like Theories}

\author{Mark P.~Bogers}
\email{mark.bogers@uis.no}
\affiliation{Department of Mathematics and Physics, University of Stavanger, 4036 Stavanger, Norway}

\author{Tom\'{a}\v{s} Brauner}
\email{tomas.brauner@uis.no}
\affiliation{Department of Mathematics and Physics, University of Stavanger, 4036 Stavanger, Norway}

\begin{abstract}
We use Lie-algebraic arguments to classify Lorentz-invariant theories of massless interacting scalars that feature coordinate-dependent redundant symmetries of the Galileon type. We show that such theories are determined, up to a set of low-energy effective couplings, by specifying an affine representation of the Lie algebra of physical, non-redundant internal symmetries and an invariant metric on its target space. This creates an infinite catalog of theories relevant for both cosmology and high-energy physics thanks to their special properties such as enhanced scaling of scattering amplitudes in the soft limit.
\begin{center}
\emph{Dedicated to Ji\v{r}\'{\i} Ho\v{s}ek on the occasion of his 75${}^\text{th}$ birthday.}\\[-3ex]
\end{center}
\end{abstract}

\maketitle

%%%%%%%%%%%%%%%%%%%%%%%%%%%%%%%%%%%%%%%%%%%%%%%%%%%%%

\emph{Introduction.}---Theories of the Galileon type have aroused immense attention in the context of modified gravity~\cite{Curtright:2012gx,*Deffayet:2013lga}. The simplest example of a Galileon field theory contains a massless scalar field $\theta$ and a single interaction term, $(\de_\mu\theta)^2\Box\theta$. It has a number of remarkable properties. First, its Lagrangian density is invariant up to a surface term under the coordinate-dependent ``Galilean'' shift, $\theta(x)\to\theta(x)+\alpha+\beta_\mu x^\mu$, where $\alpha$ and $\beta_\mu$ are constant parameters. Second, the corresponding equation of motion is of second order, implying absence of instabilities due to Ostrogradski ghosts~\cite{Horndeski:1974wa,*Fairlie:1991qe}. An attractive feature for cosmological model building is that the Galilean symmetry together with the requirement of second-order equation of motion strongly restricts the action, leaving only five possible Lagrangians in four spacetime dimensions~\cite{Nicolis:2008in} (see, however, Refs.~\cite{Lombriser:2015sxa,*Lombriser:2016yzn,Baker:2017hug,*Creminelli:2017sry,*Sakstein:2017xjx,*Ezquiaga:2017ekz} for a recent discussion of observational constraints on Galileon-like theories).

Besides cosmology, the single-field Galileon theory is also interesting for high-energy physics. Indeed, all the five possible Lagrangians can be interpreted as so-called Wess-Zumino (WZ) terms~\cite{Witten:1983tw,*DHoker:1994ti,*DHoker:1995it} of the Galilean shift symmetry~\cite{Goon:2012dy}. Consequently, they are free from radiative corrections and remain exact on the quantum level~\cite{Goon:2016ihr,Hinterbichler:2010xn}. Moreover, the five Galileon Lagrangians turn out to be related by a set of duality transformations~\cite{deRham:2013hsa,*deRham:2014lqa,*Kampf:2014rka}. Their special linear combination features a nontrivial hidden symmetry~\cite{Hinterbichler:2015pqa,*Noller:2015rea,*Novotny:2016jkh}. Finally, scattering amplitudes of the Galileon field $\theta$ vanish in the limit where the momentum of one of the participating particles goes to zero (the \emph{soft limit}). This alone follows from the fact that $\theta$ can be interpreted as a Nambu-Goldstone (NG) boson  of the shift symmetry, $\theta\to\theta+\alpha$~\cite{Weinberg:1996v2}. However, the presence of the Galilean shift symmetry makes the soft limit \emph{enhanced}: the scattering amplitudes vanish with the second power of momentum. The hidden symmetry of the special Galileon further enhances its scattering amplitudes so that they scale with the third power of momentum~\cite{Cheung:2014dqa}. These features grant the Galileon a central position in the rapidly developing branch of high-energy physics which studies scattering amplitudes of massless particles~\cite{Benincasa:2013faa,*Elvang:2013cua,*Cheung:2017pzi}. Theories whose scattering amplitudes feature an enhanced soft limit are so strongly constrained that their complete classification seems feasible~\cite{Cheung:2014dqa,Cheung:2015ota,*Padilla:2016mno,Cheung:2016drk}.

The outstanding features of the original Galileon theory~\cite{Nicolis:2008in} prompted search for its generalizations, inspired by both its potential for cosmology and the desire the understand to what extent it is uniquely determined by its properties. It has been shown that the requirement of second-order equation of motion admits a much broader class of solutions than the five WZ terms with Galilean symmetry~\cite{Deffayet:2009mn,*Deffayet:2011gz}. Some modifications of the Galileon theory, preserving its special properties in a curved spacetime, were developed in Ref.~\cite{Goon:2011qf,*Goon:2011uw}.

A vast new horizon opens by considering Galileon-like theories with \emph{multiple} massless fields. The search for such theories, motivated by applications to cosmology, was initiated in Ref.~\cite{Padilla:2010de} and continued for instance in Ref.~\cite{Padilla:2012dx,*Sivanesan:2013tba,*Allys:2016hfl}, following the equation of motion line of thought. A field-theoretic approach based on symmetry considerations was taken in Refs.~\cite{Hinterbichler:2010xn,Padilla:2010ir}, leading to some straightforward multi-flavor generalizations of the original Galileon theory. The authors of Refs.~\cite{Cheung:2014dqa,Cheung:2016drk} put forward a systematic bottom-up approach from the scattering amplitude point of view. However, their technique, based on a recursive analysis of tree-level scattering amplitudes, makes it difficult to draw general conclusions, valid for an arbitrary number of field species.

Our aim in this Letter is to classify multi-flavor Galileon-like theories from the symmetry point of view. Note that the Galilean shift symmetry $\theta(x)\to\theta(x)+\alpha+\beta_\mu x^\mu$ does not commute with the Poincar\'e group, and as such has to be spontaneously broken~\cite{Coleman:1967ad}. Yet, it does not give rise to an independent NG mode. Such symmetries are dubbed \emph{redundant} and have been studied intensively~\cite{Low:2001bw,*Watanabe:2013iia,*Hinterbichler:2014cwa,*Griffin:2014bta,*Griffin:2015hxa}. Taking the presence of a redundant symmetry as the defining property of a ``Galileon-like'' theory leads to the following question: given a set of massless scalars---NG bosons---and the associated physical internal symmetry, what additional, redundant symmetries can be consistently imposed on the system? We answer this question by solving the Lie-algebraic consistency constraints on the commutators of the symmetry generators. Furthermore, we construct three concrete classes of theories satisfying all the symmetry constraints, two of which generalize the known multi-Galileon and multi-flavor Dirac-Born-Infeld (DBI) theories. As a corollary of our work, we map the landscape of theories of multiple massless scalars whose scattering amplitudes feature an enhanced soft limit~\footnote{While there seems to be no general proof that enhancement of the soft limit \emph{requires} redundant symmetry, all currently known examples of theories with an enhanced soft limit possess one.}.

%%%%%%%%%%%%%%%%%%%%%%%%%%%%%%%%%%%%%%%%%%%%%%%%%%%%%

\emph{Geometry of redundant symmetries.}---Consider a Poincar\'e-invariant theory of one or more massless scalars (NG bosons). Its symmetry generators include the angular momentum $J_{\mu\nu}$, momentum $P_\mu$, along with the internal symmetry generators $Q_i$, which are assumed to satisfy $[P_\mu,Q_i]=0$. This condition holds for most Lorentz-scalar symmetries in physics; its violation tends to lead to nonvanishing scattering amplitudes in the soft limit~\cite{Watanabe:2014hca,*Rothstein:2017twg}. With this restriction, the commutators of all the listed generators are fixed by Lorentz invariance, except for the internal symmetry algebra, $[Q_i,Q_j]=\im\str^k_{ij}Q_k$. How can the symmetry be augmented with a set of additional, redundant generators? It is known~\cite{Ivanov:1975zq} that for the new generators to be redundant, their commutator with $P_\mu$ must be a linear combination of $Q_i$, so they must be Lorentz vectors~\footnote{In principle, the redundant generators may form a hierarchy of Lorentz vectors and higher-rank tensors, leading to multiply enhanced scattering amplitudes in the soft limit. This possibility is considered elsewhere~\cite{Bogers:2018zeg}.}. A trivial example is the linear Galilean shift, corresponding to the parameter $\beta_\mu$.

We therefore add to our theory a set of generators $K_{\mu A}$, the index $A$ indicating their multiplet structure, related in some yet unknown way to the algebra of $Q_i$. Lorentz invariance now fixes the value of $[J_{\mu\nu},K_{\lambda A}]$ and restricts the remaining commutators among the generators to
\begin{align}
\label{master}
[P_\mu,K_{\nu A}]&=\im(\aa^i_Ag_{\mu\nu}Q_i+\bb_AJ_{\mu\nu}+\cc_A\epsilon_{\mu\nu\kappa\lambda}J^{\kappa\lambda}),\\
\notag
[K_{\mu A},K_{\nu B}]&=\im(\ff_{AB}J_{\mu\nu}+\g_{AB}\epsilon_{\mu\nu\kappa\lambda}J^{\kappa\lambda}+\hh^i_{AB}g_{\mu\nu}Q_i),\\
\notag
[K_{\mu A},Q_i]&=\im(\jj_{Ai}P_\mu+\kk^B_{Ai}K_{\mu B}),
\end{align}
with as yet undetermined coefficients $\aa^i_A,\dotsc$. In order for the addition of the generators $K_{\mu A}$ to be consistent, the full Lie algebra must satisfy a set of Jacobi identities. A straightforward calculation shows that $\bb_A=\cc_A=\g_{AB}=0$. The remaining coefficients can be nonzero, but are restricted by a set of conditions that can be equivalently expressed as follows~(see the
Appendix for the list of constraints from Jacobi identities).

First, define a set of linear combinations
\begin{equation}
Q_A\equiv\aa^i_AQ_i,\qquad
Q_{AB}\equiv\hh^i_{AB}Q_i.
\label{QAdef}
\end{equation}
The commutation relations between $P_\mu$, $K_{\mu A}$, $Q_A$ and $Q_{AB}$ necessarily take the form
\begin{equation}
\begin{split}
[P_\mu,K_{\nu A}]={}&\im g_{\mu\nu} Q_A,\\
[K_{\mu A},K_{\nu B}]={}&\im(\ff_{AB}J_{\mu\nu}+g_{\mu\nu}Q_{AB}),\\
[K_{\mu A},Q_B]={}&-\im\ff_{AB}P_\mu,\\
[K_{\mu C},Q_{AB}]={}&\im(\ff_{AC}K_{\mu B}-\ff_{BC}K_{\mu A}),\\
[Q_A,Q_B]={}&0,\\
[Q_{AB},Q_C]={}&\im(\ff_{BC}Q_A-\ff_{AC}Q_B),\\
[Q_{AB},Q_{CD}]={}&\im(\ff_{AD}Q_{BC}+\ff_{BC}Q_{AD}\\
&-\ff_{AC}Q_{BD}-\ff_{BD}Q_{AC}).
\end{split}
\label{comm1}
\end{equation}
The first commutator is nothing but a multi-flavor generalization of the Galileon algebra. We therefore expect $Q_A$ to play the role of shift symmetries, acting on a space of Galileon-like fields $\theta^A$, which we will refer to as the \emph{Galileon space}. Next, introduce the set of block matrices
\begin{equation}
(T_i)^A_{\phantom AB}\equiv
\left(\begin{array}{c|c}
-\im\kk^A_{Bi} & 0\\
\hline
\jj_{Bi} & 0
\end{array}\right).
\label{comm2}
\end{equation}
These define a set of affine maps (combinations of linear transformations and translations) on a vector space, isomorphic to the Galileon space. Given that $[T_i,T_j]=\im\str^k_{ij}T_k$, the matrices $T_i$ generate an affine representation of the Lie algebra of $Q_i$. In addition, they define the adjoint action of $Q_i$ on $Q_A$ and $Q_{AB}$. By grouping the latter two as,
\begin{equation}
L_{AB}\equiv
\left(\begin{array}{c|c}
Q_{AB} & \im Q_A\\
\hline
-\im Q_B & 0
\end{array}\right),
\label{comm3}
\end{equation}
we get an object that transforms as a rank-two antisymmetric tensor under the representation $T_i$, that is,
\begin{equation}
[Q_i,L_{AB}]=(T^T_iL+LT_i)_{AB}.
\label{comm4}
\end{equation}

Altogether, the most general symmetry algebra, obtained by augmenting an internal symmetry with a set of redundant generators $K_{\mu A}$, is fully determined by: (i) the algebra of internal generators $Q_i$, (ii) its affine representation $T_i$, (iii) the matrix $\ff_{AB}$ which forms an invariant rank-two symmetric tensor of the representation $T_i$.

The last four lines in Eq.~\eqref{comm1} are a facsimile of the Poincar\'e algebra, except that they refer to the Galileon space and use the metric $\ff_{AB}$. Altogether, the commutators listed in Eq.~\eqref{comm1} together with those of the Poincar\'e group admit a remarkable geometric interpretation: they generate the group of isometries of the direct sum of the flat Minkowski spacetime and the Galileon space, equipped with the metric $g_{\mu\nu}\oplus\ff_{AB}$. The generators $Q_A$ naturally represent translations and $Q_{AB}$ rotations in the Galileon space, and $K_{\mu A}$ generate rotations between the Minkowski and the Galileon space. This picture is strongly reminiscent of the probe brane construction of DBI and Galileon actions~\cite{deRham:2010eu}. It should, however, be treated with some care: it is, for instance, not a priori guaranteed that the ``metric'' $\ff_{AB}$ is nonsingular or that all the generators $Q_{AB}$ are linearly independent, or even nonzero.

This is our main result, which is fully general, only assuming that the physical symmetry generators $Q_i$ commute with the whole Poincar\'e group and that all the redundant generators are Lorentz vectors. Although we only discussed the Lie algebra of symmetry generators so far, for spontaneously broken symmetries this fixes the action, and thus all physical observables, up to a set of low-energy effective couplings~\cite{Weinberg:1996v2}. It is necessary to clarify, though, for which of the found Lie algebras a nontrivial action in fact exists.

%%%%%%%%%%%%%%%%%%%%%%%%%%%%%%%%%%%%%%%%%%%%%%%%%%%%%

\emph{Generalized Dirac-Born-Infeld theory.}---To find an answer to the above question, let us first assume that the metric $\ff_{AB}$ is nonsingular. Its inverse, $\ff^{AB}$, must also be invariant under the representation $T_i$. The generators $Q_i$ can then be redefined as
\begin{equation}
\tilde Q_i\equiv Q_i+\jj_{Ai}\ff^{AB}Q_B;
\end{equation}
their commutators with other generators, cf.~Eq.~\eqref{comm4}, reduce to
\begin{equation}
\begin{gathered}[2]
[\tilde Q_i,K_{\mu A}]=(t_i)^B_{\phantom BA}K_{\mu B},\qquad
[\tilde Q_i,Q_A]=(t_i)^B_{\phantom BA}Q_B,\\
[\tilde Q_i,Q_{AB}]=(t_i)^C_{\phantom CA}Q_{CB}+(t_i)^C_{\phantom CB}Q_{AC},
\end{gathered}
\label{commDBI}
\end{equation}
where $(t_i)^A_{\phantom AB}\equiv-\im\kk^A_{Bi}$ is a linear representation of $Q_i$. Since $\tilde Q_A\equiv a^i_A\tilde Q_i=0$, the scalar generators split into two sets: the shift generators $Q_A$ and those of $\tilde Q_i$ that are nonzero. ($Q_{AB}$ are included among these thanks to $\tilde Q_{AB}\equiv c^i_{AB}\tilde Q_i=Q_{AB}$.) The $\tilde Q_i$ still define a closed Lie algebra, $[\tilde Q_i,\tilde Q_j]=\im\str^k_{ij}\tilde Q_k$. Owing to Eq.~\eqref{commDBI}, the symmetry algebra has the structure of a semidirect sum, where the subalgebra of $\tilde Q_i$ acts on the other generators through the representation $t_i$. In this case, the symmetry structure of the theory is thus fixed by giving: (i) the Lie algebra of $\tilde Q_i$, (ii) its representation $t_i$, (iii) the metric $\ff_{AB}$.

To proceed towards a construction of concrete actions, we use the method of nonlinear realizations, known as the coset construction~\cite{Coleman:1969sm,*Callan:1969sn}, in the version valid for spacetime symmetries~\cite{Volkov:1973vd,*Ogievetsky}. Denoting the broken generators among the $\tilde Q_i$ as $\tilde Q_a$, we parameterize the coset space as
\begin{equation}
U(x,\theta,\xi)\equiv e^{\im x^\mu P_\mu}e^{\im\theta^A Q_A}e^{\im\xi^{\mu A}K_{\mu A}}e^{\im\theta^a\tilde Q_a}.
\label{coset}
\end{equation}

Invariant actions can be built using the Maurer-Cartan (MC) form, $\omega\equiv-\im U^{-1}\dd U$. This can be decomposed in the basis of symmetry generators, the most interesting components for our purposes being $\omega^\mu_PP_\mu$ and $\omega^A_QQ_A$. Introducing the shorthand notation $\Co(x)\equiv\cosh\sqrt x$ and $\Si(x)\equiv\sinh\sqrt x/\sqrt x$, and the matrices $\XX_\mu^{\phantom\mu\nu}\equiv\ff_{AB}\xi^A_\mu\xi^{\nu B}$ and $\LL_A^{\phantom AB}\equiv\ff_{AC}\xi^{\mu C}\xi_\mu^B$, these components of the MC form can be expressed as
\begin{align}
\omega^\mu_P&=\dd x^\nu(\Co\XX)_\nu^{\phantom\nu\mu}-\dd\theta^A\ff_{AB}\xi^{\nu B}(\Si\XX)_\nu^{\phantom\nu\mu},\\
\notag
\omega^A_Q&=(e^{-\im\theta^at_a})^A_{\phantom AB}\bigl[\dd\theta^C(\Co\LL)_C^{\phantom CB}-\dd x^\mu\xi_\mu^C(\Si\LL)_C^{\phantom CB}\bigr].
\end{align}
The latter can be used to eliminate the unphysical fields $\xi^{\mu A}$ by imposing an inverse Higgs constraint (IHC)~\cite{Ivanov:1975zq,Brauner:2014aha,*Klein:2017npd}. Setting $\omega^A_Q=0$, the $\xi^{\mu A}$ are thus given implicitly by solving the algebraic equation $\de_\mu\theta^A=\xi^B_\mu(\Si\LL/\Co\LL)_B^{\phantom BA}$. The $\omega^\mu_P$ defines a covariant vielbein $e^\alpha_\mu$ by $\omega^\alpha_P\equiv e^\alpha_\mu\dd x^\mu$. This in turn leads to the metric $G_{\mu\nu}\equiv g_{\alpha\beta}e^\alpha_\mu e^\beta_\nu$. The leading-order action is obtained by integrating the invariant volume element, $\dd^4x\sqrt{-|G|}$, and upon solving the IHC becomes 
\begin{equation}
S_\text{DBI}=\int\dd^4x\sqrt{-\bigl|g_{\mu\nu}-\ff_{AB}\de_\mu\theta^A\de_\nu\theta^B\bigr|}.
\label{DBI}
\end{equation}
This is the action of the multi-flavor DBI theory as outlined e.g.~in Ref.~\cite{Cheung:2016drk}. It describes a four-dimensional brane embedded in a $(4+N)$-dimensional flat spacetime, with the induced metric $G_{\mu\nu}$ on the brane. The $N$ NG modes $\theta^A$ arise from the $N$ spontaneously broken translations, $Q_A$, in the extra dimensions. In the case of $N=1$, Eq.~\eqref{DBI} can be cast as $S_\text{DBI}=\int\dd^4x\sqrt{1-v(\de_\mu\theta)^2}$, where $v\equiv g_{A=1,B=1}$~\cite{Cheung:2014dqa}.

The multi-flavor DBI theory discussed in Ref.~\cite{Cheung:2016drk} is based on the assumption that the extra-dimensional rotations generated by $Q_{AB}$ remain unbroken, and that there are no other Lorentz-scalar symmetry generators apart from $Q_A$ and $Q_{AB}$. These assumptions are not required in our construction. The presence of other broken generators than $Q_A$ leads to additional NG modes, $\theta^a$. While these do not enter the action~\eqref{DBI}, they do appear in the general DBI-like action, taking the form
\begin{equation}
S_\text{DBI-like}=\int\dd^4x\sqrt{-|G|}\La_\text{inv}(\theta^A,\theta^a),
\label{DBINLO}
\end{equation}
where $\La_\text{inv}$ is an invariant Lagrangian density, built using the remaining components of the MC form: $\omega^{\mu\nu}_J$, $\omega^{A}_{K\mu}$, $\omega^i_{\tilde Q}$. One can e.g.~take as $\La_\text{inv}$ a Lagrangian for $\theta^a$ alone, constructed using standard methods~\cite{Callan:1969sn}, provided that indices inside are contracted using $G_{\mu\nu}$. The simplest example of such a theory corresponds to taking $N=1$ and a single, spontaneously broken generator $\tilde Q$ with the associated NG field $\phi$. The minimal interaction between $\theta$ and $\phi$ arises from the action
\begin{equation}
S_\text{DBI-like}=\int\dd^4x\sqrt{-|G|}\,G^{\mu\nu}\de_\mu\phi\de_\nu\phi,
\end{equation}
where $G_{\mu\nu}=g_{\mu\nu}-v\de_\mu\theta\de_\nu\theta$.

%%%%%%%%%%%%%%%%%%%%%%%%%%%%%%%%%%%%%%%%%%%%%%%%%%%%%

\emph{Generalized Galileon theory.}---To concretely work out another infinite class of theories that fit into our general framework, let us assume that $\jj_{Ai}=0$. This assumption is natural: by means of the Jacobi identities, it implies $\ff_{AB}=0$~(see the Appendix), and hence makes $Q_i$ and $K_{\mu A}$ form a closed Lie algebra. This is the most general Lie algebra structure where \emph{both} $Q_i$ and $K_{\mu A}$ generate an internal symmetry.

For the sake of simplicity, we will for the moment also assume that $\hh^i_{AB}=0$, that is, $Q_{AB}=0$. Such systems are then described by an extremely simple structure,
\begin{equation}
\begin{gathered}[2]
[P_\mu,K_{\nu A}]=\im g_{\mu\nu} Q_A,\\
[Q_i,K_{\mu A}]=(t_i)^B_{\phantom BA}K_{\mu B},\quad
[Q_i,Q_A]=(t_i)^B_{\phantom BA}Q_B,
\end{gathered}
\label{commGalileon}
\end{equation}
along with $[Q_A,Q_B]=[K_{\mu A},K_{\nu B}]=[K_{\mu A},Q_B]=0$. All the commutators are now determined by specifying: (i) the algebra of generators $Q_i$, (ii) its Abelian ideal generated by $Q_A$. Finally, we will \emph{assume} that, as for the DBI-like theories, all the generators $Q_i$ can be split into subsets $\tilde Q_i$ and $Q_A$ such that the $\tilde Q_i$s themselves form a closed subalgebra~\footnote{The simplest example of a Lie algebra that does \emph{not} satisfy this assumption is the Heisenberg algebra. Other examples are provided generally by Lie algebras possessing central charges.}.

The action for such theories can be obtained as above, using the parameterization~\eqref{coset} of the coset space. In this parameterization, we obtain the Galileon-like transformation rule under $Q_A$ and $K_{\mu A}$, $\theta^A\to\theta^A+\alpha^A+\beta^A_\mu x^\mu$. The generators $\tilde Q_i$, though, act on $\theta^A$ and $\xi^{\mu A}$ linearly through the matrices $t_i$. The only nontrivial pieces of the MC form are
\begin{equation}
\begin{split}
\omega^A_{K\mu}&=(e^{-\im\theta^at_a})^A_{\phantom AB}\dd\xi^B_\mu,\\
\omega^A_Q&=(e^{-\im\theta^at_a})^A_{\phantom AB}(\dd\theta^B-\xi^B_\mu\dd x^\mu),
\end{split}
\label{MCGalileon}
\end{equation}
together with the MC form for $\tilde Q_i$, $\omega^i_{\tilde Q}$.

The auxiliary fields $\xi^{\mu A}$ are eventually eliminated by using the IHC $\omega^A_Q=0$, that is, $\xi^A_\mu=\de_\mu\theta^A$. It turns out that the whole class of theories possesses a set of WZ terms, invariant under the symmetry only up to a surface term~\cite{Bogers:2018zeg},
\begin{equation}
\La^\text{WZ}_k=c_{A_1\dotsb A_k}\theta^{A_1}G_{k-1}^{A_2\dotsb A_k},
\label{multiGal}
\end{equation}
where $G_0\equiv1$ and $G_k$ for $k=1,2,3,4$ are defined by
\begin{equation}
\begin{split}
G^{A_1\dotsb A_k}_k\equiv{}&\frac1{(4-k)!}\epsilon_{\alpha_1\dotsb\alpha_k\mu_{k+1}\dotsb\mu_4}\epsilon^{\beta_1\dotsb\beta_k\mu_{k+1}\dotsb\mu_4}\\
&\times(\de_{\beta_1}\de^{\alpha_1}\theta^{A_1})\dotsb(\de_{\beta_k}\de^{\alpha_k}\theta^{A_k});
\end{split}
\end{equation}
$c_{A_1\dotsb A_k}$ must be fully symmetric invariant tensors of the representation $t_i$. Existing multi-Galileon Lagrangians~\cite{Goon:2012dy} correspond to the special case where the generators $\tilde Q_i$ are unbroken, forming a compact Lie algebra such as $\gr{SO}(N)$ or $\gr{SU}(N)$. For an $\gr{SO}(N)$ vector of fields $\theta^A$, for instance, only $\La^\text{WZ}_2$ and $\La^\text{WZ}_4$ exist, for which respectively $c_{AB}=\delta_{AB}$ and $c_{ABCD}=\delta_{AB}\delta_{CD}+\delta_{AC}\delta_{BD}+\delta_{AD}\delta_{BC}$. Eq.~\eqref{multiGal} is fully general in that it applies to an arbitrary Lie algebra, generated by $\tilde Q_i$, and its arbitrary real (finite-dimensional) representation $t_i$, and allows for $\tilde Q_i$ to be spontaneously broken.

As for the DBI theory~\eqref{DBI}, the WZ terms~\eqref{multiGal} are blind to the $\theta^a$ fields, regardless of the symmetry-breaking pattern. Interactions between the Galileon and non-Galileon sector enter only through the strictly invariant part of the Lagrangian, and can be constructed by taking a product of $\omega_{K\mu}^A$ and $\omega_{\tilde Q}^a$ or their derivatives. These invariant Lagrangians realize the enhanced soft limit of scattering amplitudes of the Galileon modes trivially in that they only contain operators with at least two derivatives on each $\theta^A$.

%%%%%%%%%%%%%%%%%%%%%%%%%%%%%%%%%%%%%%%%%%%%%%%%%%%%%

\emph{Twisted Galileon theory.}---We shall now restore the generators $Q_{AB}$, otherwise keeping the same assumptions as in the construction of the generalized Galileon theory. The resulting Lie algebra features a twisted commutator,
\begin{equation}
[K_{\mu A},K_{\nu B}]=\im g_{\mu\nu}Q_{AB}.
\label{commtwist}
\end{equation}
Eq.~\eqref{commGalileon} still holds, and $Q_{AB}$ transforms as a rank-two tensor under the adjoint action of $Q_i$. All the generators $Q_A$ and $Q_{AB}$ commute with each other and with $K_{\mu C}$.

The coset construction proceeds as above upon the replacement $e^{\im\theta^A Q_A}\to e^{\im\theta^A Q_A}e^{\frac\im2\theta^{AB}Q_{AB}}$ in Eq.~\eqref{coset}, where $\theta^{AB}$ are NG fields for those of the $Q_{AB}$ generators that are spontaneously broken. All our conclusions reached in the discussion of the generalized Galileon theory---Eq.~\eqref{MCGalileon} and below---remain valid. However, the new component of the MC form,
\begin{equation}
\begin{split}
\omega^{AB}_Q={}&(e^{-\im\theta^at_a})^A_{\phantom AC}(e^{-\im\theta^bt_b})^B_{\phantom BD}\\
&\times\bigl[\dd\theta^{CD}+\tfrac12\bigl(\xi^C_\mu\dd\xi^{\mu D}-\xi^D_\mu\dd\xi^{\mu C}\bigr)\bigr],
\end{split}
\end{equation}
automatically induces interactions of the Galileon modes $\theta^A$ with the non-Galileon modes $\theta^{AB}$. The entanglement of the $\theta^A$ and $\theta^{AB}$ sectors is due to the transformation of $\theta^{AB}$ under $K_{\mu A}$, which reads $\theta^{AB}\to\theta^{AB}-\tfrac12(\beta^A_\mu\xi^{\mu B}-\beta^B_\mu\xi^{\mu A})$. Upon imposing the IHC, $\xi_\mu^A=\de_\mu\theta^A$, a Lagrangian built up from $\omega_Q^{AB}$ will contain less than two derivatives per $\theta^A$.

The simplest example of such a theory is obtained by taking $\gr{SO(2)}$ for the algebra of $\tilde Q_i$ and assuming that it remains unbroken. Then, $Q_A$ reduces to an $\gr{SO(2)}$ vector $Q_1,Q_2$, and $Q_{AB}$ to an $\gr{SO(2)}$ singlet $Q_{12}$. The resulting interaction Lagrangian, complementing the pure Galileon WZ terms, reads
\begin{equation}
\La_\text{twist}=\frac12\bigl[\de_\mu\theta^{12}+\tfrac12\bigl(\de^\nu\theta^1\de_\mu\de_\nu\theta^2-\de^\nu\theta^2\de_\mu\de_\nu\theta^1\bigr)\bigr]^2.
\label{Ltwisted}
\end{equation}
Interestingly, the scattering amplitudes of $\theta^1$, $\theta^2$ in this theory do \emph{not} have an enhanced soft limit in spite of the Galilean shift symmetry, acting on these fields. This can be attributed to the cubic part of the Lagrangian~\eqref{Ltwisted}, which leads to collinear singularities in the soft limit~\cite{Cheung:2014dqa}. We expect the same behavior for all twisted Galileon theories.

%%%%%%%%%%%%%%%%%%%%%%%%%%%%%%%%%%%%%%%%%%%%%%%%%%%%%

\emph{Conclusions.}---Under only very mild assumptions, we have found the most general symmetry structure, admitting a set of generators which, though spontaneously broken, do not give rise to NG bosons. We thereby mapped the landscape of possible theories of multiple massless scalars whose scattering amplitudes exhibit an enhanced soft limit~\cite{Cheung:2014dqa}. Our results reduce the construction of such theories to the problem of finding an affine representation of the Lie algebra of physical internal symmetries and an invariant metric on its target space.

Intriguingly, our analysis singles out the multi-Galileon and multi-flavor DBI theories as possibly the only systems where the scattering amplitudes of \emph{all} the modes have an enhanced soft limit. We stress, however, that our discussion of the solutions to the Lie-algebraic constraints was not exhaustive. The gap for some options not considered here is narrowed down, or closed, in the Appendix. This includes systems where the metric $\ff_{AB}$ is singular but nonzero, algebras of the type~\eqref{commGalileon} where the non-Galileon generators $\tilde Q_i$ do not form a closed subalgebra, and algebras of the type~\eqref{commtwist} where the generators $Q_{AB}$ are not linearly independent of $Q_A$.

Finally, our analysis was restricted to Lorentz-invariant theories, yet our Lie-algebraic approach can be applied equally well to nonrelativistic systems. In condensed-matter physics, there is a plethora of naturally occurring redundant symmetries such as Galilei boosts or rotations in crystalline solids. The need to understand the consequences of such symmetries provides a strong motivation for an extension of the present results, which will be addressed in our future work.

%%%%%%%%%%%%%%%%%%%%%%%%%%%%%%%%%%%%%%%%%%%%%%%%%%%%%

\emph{Acknowledgments.}---We thank Torsten Schoeneberg and Qiaochu Yuan for advice regarding the structure of Lie algebras possessing an Abelian ideal. This work has been supported by a ToppForsk-UiS grant no.~PR-10614.

%%%%%%%%%%%%%%%%%%%%%%%%%%%%%%%%%%%%%%%%%%%%%%%%%%%%%

\section*{Appendix}

\subsection*{Constraints on the symmetry algebra from Jacobi identities}

Here we provide some details behind the geometric picture of symmetry algebras with redundant generators, developed in the main text. We take as our starting point the ansatz for commutation relations of the generators $J_{\mu\nu}$, $P_\mu$, $K_{\mu A}$, $Q_i$ in Eq.~(\ref{master}) of the main text. The commutators of a given Lie algebra must satisfy, apart from bilinearity and antisymmetry, the Jacobi identity
\begin{equation}
[X,[Y,Z]]+[Y,[Z,X]]+[Z,[X,Y]]=0
\end{equation}
for any choice of generators $X$, $Y$, $Z$. By inserting in turn all possible triples of generators, a straightforward, albeit somewhat tedious, calculation leads to $\bb_A=\cc_A=\g_{AB}=0$ as well as the following set of nonlinear constraints,
\begin{align}
\label{eq12}
\kk^C_{Ai}\kk^B_{Cj}-\kk^C_{Aj}\kk^B_{Ci}&=\str^k_{ij}\kk^B_{Ak},\\
\label{eq13}
\aa^i_A\kk^B_{Ci}&=0,\\
\label{eq14}
\kk^B_{Ai}\aa^k_B+\aa^j_A\str^k_{ij}&=0,\\
\label{eq15}
\kk^B_{Ai}\jj_{Bj}-\kk^B_{Aj}\jj_{Bi}&=\str^k_{ij}\jj_{Ak},\\
\label{eq16}
\ff_{AB}&=-\aa^i_A\jj_{Bi},\\
\label{eq17}
\ff_{AB}&=-\aa^i_B\jj_{Ai},\\
\label{eq18}
\kk^C_{Bi}\ff_{AC}+\kk^C_{Ai}\ff_{CB}&=0,\\
\label{eq19}
\ff_{AC}\delta^D_B-\ff_{BC}\delta^D_A&=\hh^i_{AB}\kk^D_{Ci},\\
\label{eq20}
\hh^i_{AB}\jj_{Ci}&=0,\\
\label{eq21}
\kk^C_{Ai}\hh^k_{CB}+\kk^C_{Bi}\hh^k_{AC}-\aa^k_A\jj_{Bi}+\aa^k_B\jj_{Ai}&=-\str^k_{ij}\hh^j_{AB}.
\end{align}

Some of these constraints have an obvious interpretation. For instance, Eq.~\eqref{eq12} guarantees that the matrices $(t_i)^A_{\phantom AB}\equiv-\im\kk^A_{Bi}$ define a representation of the algebra of the internal generators $Q_i$ with the structure constants $\str^k_{ij}$. By Eq.~\eqref{eq13}, the generators $Q_A\equiv\aa^i_AQ_i$ then vanish in this representation. Eq.~\eqref{eq14} is equivalent to the relation
\begin{equation}
[Q_i,Q_A]=(t_i)^B_{\phantom BA}Q_B.
\label{eq22}
\end{equation}
Together, the latter two conditions ensure that
\begin{equation}
[Q_A,Q_B]=0.
\end{equation}
Next, Eq.~\eqref{eq21} is equivalent to the commutator
\begin{equation}
[Q_i,Q_{AB}]=(t_i)^C_{\phantom CA}Q_{CB}+(t_i)^C_{\phantom CB}Q_{AC}+\im(\jj_{Bi}Q_A-\jj_{Ai}Q_B),
\label{eq24}
\end{equation}
where $Q^i_{AB}\equiv\hh^i_{AB}Q_i$. Eq.~\eqref{eq22} and~\eqref{eq24} together are then encoded in Eq.~(\ref{comm4}) of the main text. Likewise, Eqs.~\eqref{eq13},~\eqref{eq16} and~\eqref{eq17} together are equivalent to the constraint
\begin{equation}
[K_{\mu A},Q_B]=-\im\ff_{AB}P_\mu
\end{equation}
with symmetric $\ff_{AB}$. Eqs.~\eqref{eq19} and~\eqref{eq20} together are equivalent to
\begin{equation}
[K_{\mu C},Q_{AB}]=\im(\ff_{AC}K_{\mu B}-\ff_{BC}K_{\mu A}).
\end{equation}
Eq.~\eqref{eq15} is required to ensure (and follows from) that the matrices $T_i$, defined in Eq.~(\ref{comm2}) of the main text, form a representation of the Lie algebra of $Q_i$. Finally, Eq.~\eqref{eq18} guarantees that the matrix $\ff_{AB}$ is an invariant tensor of the representation $t_i$. This is, in fact, not an independent constraint: it follows as a necessary consequence from Eqs.~\eqref{eq13}--\eqref{eq16}.

This concludes the proof that all the constraints following from the Jacobi identities can be obtained from the commutation relations, listed in Eqs.~(\ref{comm1})--(\ref{comm4}) of the main text. That the opposite implication also holds will be asserted once we have proven the last two commutators in Eq.~(\ref{comm1}) of the main text. The first of them is obtained by multiplying Eq.~\eqref{eq22} by $\hh^i_{BC}$ and using Eq.~\eqref{eq19}. The second commutator then follows analogously from Eq.~\eqref{eq24} with additional use of Eq.~\eqref{eq20}.

Altogether, we have shown that Eqs.~(\ref{QAdef})--(\ref{comm4}) in the main  text are an equivalent representation of the set of constraints following from the Jacobi identities for the generators of the symmetry algebra.

%%%%%%%%%%%%%%%%%%%%%%%%%%%%%%%%%%%%%%%%%%%%%%%%%%%%%

\subsection*{Galileon Lagrangians from the DBI theory}

The multi-Galileon Lie algebra, given in Eq.~(\ref{commGalileon}) of the main text, can be obtained from the multi-flavor DBI algebra by setting $Q_{AB}=0$ and going to the limit $\ff_{AB}\to0$. This raises the possibility that also the multi-Galileon Lagrangians can be obtained by taking a similar limit from their DBI counterparts~(see also Ref.~\cite{deRham:2010eu}). For instance, the leading-order DBI action, $\int\dd^4x\sqrt{-|G|}$, when expanded in powers of $\ff_{AB}$, yields
\begin{equation}
S=\int\dd^4x\left(1-\frac12\ff_{AB}\de_\mu\theta^A\de^\mu\theta^B\right)+\mathcal O(\ff^2).
\label{DBIlimit}
\end{equation}
The $\mathcal O(\ff^1)$ piece herein is precisely the second Galileon Lagrangian which provides a kinetic term for the fields $\theta^A$.

To understand \emph{why} this procedure leads to the Galileon Lagrangians, which after all are WZ terms indicating a nontrivial realization of the Galilean shift symmetry, let us have a look at the transformation properties of the fields. The transformation rules are obtained from the coset parameterization, Eq.~(\ref{coset}) of the main text. While the spacetime translations and the internal transformations generated by $Q_A$ act trivially as a shift of $x^\mu$ and $\theta^A$, respectively, the transformations generated by $K_{\mu A}$ are essential. Multiplying $U$ from the left by $e^{\im\beta^{\mu A}K_{\mu A}}$ we obtain, after some algebra, the rules
\begin{equation}
\begin{split}
x^\mu&\to x^\nu(\Co\XX_\beta)_\nu^{\phantom\nu\mu}+\theta^A\ff_{AB}\beta^{\nu B}(\Si\XX_\beta)_\nu^{\phantom\nu\mu},\\
\theta^A&\to\theta^B(\Co\LL_\beta)_B^{\phantom BA}+x^\mu\beta^B_\mu(\Si\LL_\beta)_B^{\phantom BA},
\end{split}
\end{equation}
where $(\XX_\beta)_\mu^{\phantom\mu\nu}\equiv\ff_{AB}\beta^A_\mu\beta^{\nu B}$ and $(\LL_\beta)_A^{\phantom AB}\equiv\ff_{AC}\beta^{\mu C}\beta^B_\mu$. In fact, an infinitesimal version of these transformation rules is sufficient for our purposes, and easier to work with,
\begin{equation}
\begin{split}
x^\mu&\to x^\mu+\theta^A\ff_{AB}\beta^{\mu B}+\mathcal O(g\beta^2),\\
\theta^A&\to\theta^A+x^\mu\beta^A_\mu+\mathcal O(g\beta^2).
\end{split}
\label{DBItransfo}
\end{equation}
In the limit $\ff_{AB}\to0$, these naturally reproduce the transformation under the $K_{\mu A}$ generator of the Galileon algebra.

Now the $\mathcal O(g^0)$ piece of the DBI action~\eqref{DBIlimit} changes by a mere Jacobian, $\dd^4x\to\dd^4x(1+\ff_{AB}\de_\mu\theta^A\beta^{\mu B}+\dotsb)$. This is exactly what is needed to cancel the surface term that arises from the transformation of the $\mathcal O(g^1)$ part of the action. A similar argument relates the other Galileon Lagrangians to the higher-order DBI actions. We can thus understand the origin of the Galileon Lagrangians and their WZ nature by starting from the DBI algebra and action and performing the appropriate contraction on them.

Let us now see if genuinely new theories can be constructed for which $\ff_{AB}$ is nonzero yet not invertible. Rather than trying to solve all the Lie-algebraic constraints in the full generality, we will again take the multi-flavor DBI theory as the starting point and then take the limit in which \emph{some} of the components of the metric $\ff_{AB}$ vanish.

First of all, $\ff_{AB}$ can be assumed diagonal without loss of generality. (Being real and symmetric, it can always be diagonalized by a change of basis of the generators $Q_A$.) We now want to send some of the eigenvalues of $\ff_{AB}$ to zero, while keeping the others fixed. Let us for the sake of simplicity consider the simplest case of two NG fields $\theta^A$ with
\begin{equation}
\ff_{AB}=\begin{pmatrix}
1 & 0\\0 & \epsilon
\end{pmatrix}.
\end{equation}
The infinitesimal symmetry transformation~\eqref{DBItransfo} becomes
\begin{equation}
\begin{split}
x^\mu&\to x^\mu+\theta^1\beta^{\mu1}+\epsilon\theta^2\beta^{\mu2}+\dotsb,\\
\theta^A&\to\theta^A+x^\mu\beta^A_\mu+\dotsb,
\end{split}
\label{inftytransfo}
\end{equation}
which in the limit $\epsilon\to0$ corresponds to a DBI-like symmetry acting on $\theta^1$ and a Galileon-like symmetry acting on $\theta^2$. The DBI action $\int\dd^4x\sqrt{-|G_\epsilon|}$, expanded in $\epsilon$, leads to
\begin{align}
\notag
S&=\int\dd^4x\sqrt{1-(\de_\mu\theta^1)^2}\left(1-\frac\epsilon2G_0^{\mu\nu}\de_\mu\theta^2\de_\nu\theta^2\right)+\mathcal O(\epsilon^2)\\
&\equiv S_0+\epsilon S_1+\epsilon^2S_2+\dotsb,
\end{align}
where
\begin{equation}
(G_\epsilon)_{\mu\nu}\equiv g_{\mu\nu}-\de_\mu\theta^1\de_\nu\theta^1-\epsilon\de_\mu\theta^2\de_\nu\theta^2.
\end{equation}
The variation of the full action under the symmetry transformation~\eqref{inftytransfo} must vanish order by order in $\epsilon$. This means in particular that the $S_0$ piece must be invariant in the $\epsilon\to0$ limit, whereas the $\mathcal O(\epsilon^1)$ part of the variation of $S_0$ must cancel the $\epsilon\to0$ limit of the variation of $S_1$, and so on. However, it is easy to check that the $\mathcal O(\epsilon^1)$ variation of $S_0$ is \emph{not} a surface term anymore, owing to the fact that it is not given merely by the Jacobian of the coordinate transformation. Consequently, $S_1$ is no longer invariant on its own.

An explicit inversion of the metric $(G_\epsilon)_{\mu\nu}$ gives
\begin{equation}
S_1=-\frac12\int\dd^4x\left[(\de_\mu\theta^2)^2\sqrt{1-(\de_\nu\theta^1)^2}+\frac{(\de_\mu\theta^1\de^\mu\theta^2)^2}{\sqrt{1-(\de_\nu\theta^1)^2}}\right].
\end{equation}
This describes a coupling of the DBI scalar $\theta^1$ to a second NG mode, $\theta^2$. However, the latter does not have the features of the Galileon: the action is not invariant under its shift linear in the coordinate, and its scattering amplitudes only vanish with the first power of momentum; their soft limit is not enhanced.

Although we have only worked out explicitly a very simple example, the same argument can clearly be applied to the general case of $\ff_{AB}$ having both zero and nonzero eigenvalues. We therefore conclude that while the multi-flavor DBI and the multi-Galileon theory are robust solutions of the Lie-algebraic constraints~\eqref{eq12}--\eqref{eq21}, a naive attempt at constructing a ``mixed'' system, interpolating between the two limits, fails. Such systems, if possible at all, would require a more thorough analysis.

%%%%%%%%%%%%%%%%%%%%%%%%%%%%%%%%%%%%%%%%%%%%%%%%%%%%%

\subsection*{Galileon algebras with central extension}

In the discussion of the Galileon-like algebras, Eq.~(\ref{commGalileon}) of the main text, we assumed that the scalar generators can be split up into subsets $\tilde Q_i$ and $Q_A$ such that the $\tilde Q_i$s themselves form a closed subalgebra. In fact, there is an infinite class of Lie algebras that do \emph{not} satisfy this assumption. The simplest example of such algebras, which we will now discuss in some detail, occurs when the representation $t_i$ is trivial. The algebra of the scalar generators then reads
\begin{equation}
[\tilde Q_i,\tilde Q_j]=\im f^k_{ij}\tilde Q_k+\im f^A_{ij}Q_A,\quad
[\tilde Q_i,Q_A]=[Q_A,Q_B]=0,
\label{central}
\end{equation}
where $\tilde Q_i$ is a maximal subset of $Q_i$ that is linearly independent of $Q_A$. Algebras of this type can be thought of as central extensions of the algebra of the $\tilde Q_i$s alone, where the $Q_A$s play the role of central charges. They can be fully classified by the second cohomology of the $\tilde Q_i$ algebra, see~e.g.~Ref.~\cite{Watanabe:2014fva}.

We can now imagine carrying out the coset construction as in the main text, using Eq.~(\ref{coset}) therein as the coset parameterization. This leads to the MC form that serves as the basic building block for construction of invariant actions. The components of the MC form satisfy a set of so-called MC equations, whose form only depends on the commutation relations among the generators,
\begin{equation}
\begin{split}
\dd\omega^A_Q&=\omega_P^\mu\wedge\omega^A_{K\mu}+\frac12f^A_{jk}\omega_{\tilde Q}^j\wedge\omega_{\tilde Q}^k,\\
\dd\omega_{\tilde Q}^i&=\frac12f^i_{jk}\omega_{\tilde Q}^j\wedge\omega_{\tilde Q}^k,
\end{split}
\end{equation}
along with $\dd\omega_P^\mu=\dd\omega^A_{K\mu}=0$.

At the end of the day, $\omega_Q^A$ is eliminated by imposing the IHC, upon which $\omega^A_{K\mu}=\dd\xi^A_\mu$ only depends on a second derivative of $\theta^A$. It is therefore not possible to write down an invariant Lagrangian that would include a kinetic term for the $\theta^A$ fields. For systems that, unlike the DBI theories, have a trivial vielbein, the kinetic term for $\theta^A$ \emph{has to} come from some WZ term. This imposes strong constraints on the existence of perturbatively well-defined theories for the $\theta^A$ fields.

In four spacetime dimensions, the WZ terms are obtained from closed invariant 5-forms that belong to the cohomology of the coset space of the broken symmetry. In case of algebras of the type~\eqref{central} with $f^A_{ij}=0$, there are five linearly independent 5-forms that are closed and invariant,
\begin{equation}
\begin{split}
\omega^1_5&=\epsilon_{\kappa\lambda\mu\nu}c_A\omega^A_Q\wedge\dd x^\kappa\wedge\dd x^\lambda\wedge\dd x^\mu\wedge\dd x^\nu,\\
\omega^2_5&=\epsilon_{\kappa\lambda\mu\nu}c_{AB}\omega^A_Q\wedge\omega^{B\kappa}_K\wedge\dd x^\lambda\wedge\dd x^\mu\wedge\dd x^\nu,\\
\omega^3_5&=\epsilon_{\kappa\lambda\mu\nu}c_{ABC}\omega^A_Q\wedge\omega^{B\kappa}_K\wedge\omega^{C\lambda}_K\wedge\dd x^\mu\wedge\dd x^\nu,\\
\omega^4_5&=\epsilon_{\kappa\lambda\mu\nu}c_{ABCD}\omega^A_Q\wedge\omega^{B\kappa}_K\wedge\omega^{C\lambda}_K\wedge\omega^{D\mu}_K\wedge\dd x^\nu,\\
\omega^5_5&=\epsilon_{\kappa\lambda\mu\nu}c_{ABCDE}\omega^A_Q\wedge\omega^{B\kappa}_K\wedge\omega^{C\lambda}_K\wedge\omega^{D\mu}_K\wedge\omega^{E\nu}_K.
\end{split}
\label{gal5forms}
\end{equation}
These are in one-to-one correspondence to the five different multi-flavor Galileon terms in four spacetime dimensions, see Ref.~\cite{Goon:2012dy}. Since both $\omega_Q^A$ and $\omega_{K\mu}^A$ contain one factor of $\theta^A$, the kinetic term for the $\theta^A$ fields comes from $\omega^2_5$.

With $f^A_{ij}\neq0$, these forms are no longer closed. We can, however, hope to restore their closedness by adding analogous 5-forms with $\omega_{\tilde Q}^i$ in place of $\omega_Q^A$, so that the contributions proportional to $\omega_{\tilde Q}^j\wedge\omega_{\tilde Q}^k$ coming from $\dd\omega_Q^A$ and $\dd\omega_{\tilde Q}^i$ cancel each other. Let us see how this could be used to construct a kinetic term for $\theta^A$. To that end, we use the 5-form
\begin{equation}
\tilde\omega^2_5=\epsilon_{\kappa\lambda\mu\nu}(c_{AB}\omega_Q^A+c_{iB}\omega_{\tilde Q}^i)\wedge\omega^{B\kappa}_K\wedge\dd x^\lambda\wedge\dd x^\mu\wedge\dd x^\nu.
\end{equation}
This form is closed if and only if the following condition is satisfied,
\begin{equation}
c_{AB}f^A_{jk}+c_{iB}f^i_{jk}=0.
\end{equation}
$c_{AB}$ must be non-singular in order to give a kinetic term for \emph{all} the fields $\theta^A$, hence we can use it as a metric to raise and lower indices. The above condition can then be solved for $f^A_{ij}$,
\begin{equation}
f^A_{ij}=-c_k^{\phantom kA}f^k_{ij}.
\end{equation}
Upon a redefinition of the generators,
\begin{equation}
\tilde Q'_i\equiv\tilde Q_i-c_i^{\phantom iA}Q_A,
\end{equation}
we then find that $[\tilde Q'_i,\tilde Q'_j]=\im f^k_{ij}\tilde Q'_k$. In other words, a WZ term giving a kinetic term for the $\theta^A$ fields only exists if the central extension of the algebra of $\tilde Q_i$ is trivial. Nontrivial central extensions represent an obstruction that rules out the existence of a perturbatively well-defined field theory.

%%%%%%%%%%%%%%%%%%%%%%%%%%%%%%%%%%%%%%%%%%%%%%%%%%%%%

\subsection*{Contracted twisted Galileon algebras}

In the construction of the twisted Galileon theories, based on Eq.~(\ref{commtwist}) of the main text, we implicitly assumed that the generator $Q_{AB}$ appearing on the right-hand side therein is linearly independent of the Galilean shift generators $Q_A$. That is, however, not necessary. By giving up this assumption, we might hope to construct genuinely new theories that do not contain any other massless scalars than the Galileon modes, whose scattering amplitudes are guaranteed to have an enhanced soft limit.

Let us therefore assume that
\begin{equation}
Q_{AB}=\lambda_{AB}^{\phantom{AB}C}Q_C,
\label{contraction}
\end{equation}
where $\lambda_{AB}^{\phantom{AB}C}$ is a set of a priori undetermined coefficients. We can think of this as a contraction of the twisted Galileon algebra that does not contain any additional scalar generators. In order for this to be consistent with Eqs.~\eqref{eq22} and~\eqref{eq24} (with $\jj_{Ai}=0$), the coefficients $\lambda_{AB}^{\phantom{AB}C}$ must satisfy the condition
\begin{equation}
(t_i)^D_{\phantom DA}\lambda_{DB}^{\phantom{DB}C}+(t_i)^D_{\phantom DB}\lambda_{AD}^{\phantom{AD}C}-(t_i)^C_{\phantom CD}\lambda_{AB}^{\phantom{AB}D}=0,
\end{equation}
which expresses the invariance of $\lambda_{AB}^{\phantom{AB}C}$ under the representation $t_i$ of the internal symmetry.

The coset construction now proceeds as in the main text, resulting in a modification of the MC form, Eq.~(\ref{MCGalileon}) therein,
\begin{equation}
\omega_Q^A=(e^{-\im\theta^at_a})^A_{\phantom AB}\bigl(\dd\theta^B-\xi^B_\mu\dd x^\mu+\tfrac12\lambda_{CD}^{\phantom{CD}B}\xi^C_\mu\dd\xi^{\mu D}\bigr);
\end{equation}
the other components of the MC form remain unchanged. This will at the end of the day lead to a rather involved IHC,
\begin{equation}
\xi^A_\mu=\de_\mu\theta^A+\frac12\lambda_{BC}^{\phantom{BC}A}\xi_\nu^B\de_\mu\xi^{\nu C},
\label{twistedIHC}
\end{equation}
which fixes $\xi^A_\mu$ in terms of $\theta^A$ through the solution of a nonlinear differential equation. The invariance of this IHC is guaranteed by a modified transformation of $\theta^A$ under $K_{\mu A}$,
\begin{equation}
\theta^A\to\theta^A+\beta_\mu^A x^\mu-\frac12\lambda_{BC}^{\phantom{BC}A}\beta_\mu^B\xi^{\mu C},
\end{equation}
along with the unchanged $\xi^A_\mu\to\xi^A_\mu+\beta^A_\mu$.

This is not the only complication we have to face. The MC equations now take the form
\begin{align}
\notag
\dd\omega_{K\mu}^A&=-\im\Omega^A_{\phantom AB}\wedge\omega_{K\mu}^B,\\
\notag
\dd\omega_Q^A&=\omega_P^\mu\wedge\omega^A_{K\mu}-\im\Omega^A_{\phantom AB}\wedge\omega_Q^B+\frac12\lambda_{BC}^{\phantom{BC}A}\omega_{K\mu}^B\wedge\omega_K^{\mu C},\\
\dd\omega_{\tilde Q}^i&=\frac12f^i_{jk}\omega_{\tilde Q}^j\wedge\omega_{\tilde Q}^k,
\end{align}
where $\Omega^A_{\phantom AB}\equiv\omega_{\tilde Q}^i(t_i)^A_{\phantom AB}$. The Galileon 5-forms~\eqref{gal5forms} will no longer be closed for generic $\lambda_{AB}^{\phantom{AB}C}$, and it is not possible to restore their closedness by adding terms proportional to $\omega_{\tilde Q}^i$. Imposing the closedness of the Galileon forms as a constraint on $\lambda_{AB}^{\phantom{AB}C}$ does not seem to have any solution with nonzero $\lambda_{AB}^{\phantom{AB}C}$ except for one spacetime dimension. There, the 2-form
\begin{equation}
\omega^2_2=c_{AB}\omega_Q^A\wedge\omega_K^B
\end{equation}
is invariant and closed provided $c_{AB}$ is a symmetric rank-two invariant tensor and $c_{D[A}\lambda_{BC]}^{\phantom{BC}D}=0$. This is solved for instance by setting $\lambda_{AB}^{\phantom{AB}C}=\delta^C_A-\delta^C_B$. It is, however, still nontrivial to solve the IHC~\eqref{twistedIHC} in order to give an explicit, local form of the Lagrangian in terms of $\theta^A$ alone.

We conclude that contracting the twisted Galileon algebra via Eq.~\eqref{contraction} is unlikely to lead to physically interesting, perturbatively well-defined theories in four spacetime dimensions.

%%%%%%%%%%%%%%%%%%%%%%%%%%%%%%%%%%%%%%%%%%%%%%%%%%%%%

\bibliography{references}

%%%%%%%%%%%%%%%%%%%%%%%%%%%%%%%%%%%%%%%%%%%%%%%%%%%%%

\end{document}